\documentclass[english]{revtex4-1}
\usepackage[T1]{fontenc}
\usepackage[latin9]{inputenc}
\usepackage{amsmath}
\usepackage{amssymb}
\usepackage{graphicx}
\usepackage{wasysym}

\makeatletter
\usepackage{babel}

\makeatother

\usepackage{babel}
\begin{document}

\title{Relativistic probability amplitudes I. Massive particles of any spin}

\author{Scott E. Hoffmann}

\address{School of Mathematics and Physics,~~\\
 The University of Queensland,~~\\
 Brisbane, QLD 4072~~\\
 Australia}
\email{scott.hoffmann@uqconnect.edu.au}

\selectlanguage{english}%
\begin{abstract}
We consider a massive particle of arbitrary spin and the basis vectors
that carry the unitary, irreducible representations of the Poincaré
group. From the complex coefficients in normalizable superpositions
of these basis vectors, we identify momentum/spin-component probability
amplitudes with the same interpretation as in the nonrelativistic
theory. We find the relativistic transformations of these amplitudes,
which are unitary in that they preserve the modulus-squared of scalar
products from frame to frame. Space inversion and time reversal are
also treated. We reconsider the Newton-Wigner construction of eigenvectors
of position and the position operator. Position/spin-component probability
amplitudes are also identified and their relativistic, unitary, transformations
derived. Again, space inversion and time reversal are considered.
For reference, we show how to construct positive energy solutions
of the Klein-Gordon and Dirac equations in terms of probability amplitudes.
We find the boost transformation of the position operator in the spinless
case and present some results on the relativity of position measurements.
We consider issues surrounding the classical concept of causality
as it applies in quantum mechanics. We briefly examine the relevance
of the results presented here for theories of interaction.
\end{abstract}
\maketitle

\section{Introduction}

This paper will deal entirely with free particles, or particles isolated
far from other particles with which they might interact. While we
will not construct a relativistic theory of interaction here, we will
see that an understanding of the behaviour of free particles is an
essential addition to any such theory.

Historically, the inclusion of the principles of relativity into quantum
mechanics has met with serious problems, with the Klein-Gordon equation
as our first example \cite{Horwitz2015,Itzykson1980}. There, it was
assumed that a spinless particle should be described by an amplitude
that transforms relativistically as a scalar function, $\varphi(x),$
of position and time ($x^{\mu}=(t,\boldsymbol{x})^{\mu})$. A manifestly
covariant equation (in terms of the four-vector derivative $\partial_{\mu}=\partial/\partial x^{\mu}$)
was proposed to determine the time evolution. This is the Klein-Gordon
equation for mass $m_{0},$
\begin{equation}
\{\partial^{\mu}\partial_{\mu}+m_{0}^{2}\}\varphi(x)=0.\label{eq:1.1}
\end{equation}

Problems arose immediately. The equation is second order in the time
derivative, unlike the nonrelativistic Schrödinger equation, changing
the nature of the initial value problem. The interpretation of the
amplitude, $\varphi(x),$ is unclear. It was evident that its modulus-squared,
$|\varphi(x)|^{2},$ could not be interpreted as a probability density,
as in the nonrelativistic theory. It does not obey an invariant conservation
equation. Then a locally conserved four-vector current density, $J_{\mathrm{KG}}^{\mu}(x)=i\{\varphi^{*}(x)\partial^{\mu}\varphi(x)-\partial^{\mu}\varphi^{*}(x)\varphi(x)\}$
was constructed, to be proposed as a probability current density.
But the zero component is not positive definite.

In addition, the Klein-Gordon equation has unphysical negative energy
solutions as well as positive energy solutions. This should be obvious,
since the equation is expressing ``$p^{2}=m_{0}^{2}$'' instead
of ``$p^{0}=+\sqrt{\boldsymbol{p}^{2}+m_{0}^{2}}$''. A particle
with a negative energy has a velocity in the opposite direction to
its momentum. Allowing the inclusion of negative as well as positive
energies in the one superposition leads to rapid, real oscillations
of both momentum and position amplitudes, destroying any possibility
of normalization.

The four-component Dirac amplitude \cite{Horwitz2015,Itzykson1980}
for spin-$1/2$ electrons and positrons also has an interpretation
problem. The index of the four components is not the eigenvalue of
an observable and the values of the components change when the gamma
matrix representation is changed. The triumph of Dirac's theory, though,
is that it constructs a locally transforming four-vector current density
with negative/positive definite zero component for electrons/positrons.
The four-component amplitude is perhaps best seen as a necessary intermediate
step in this construction. As with the Klein-Gordon equation, the
Dirac equation allows unphysical negative energy solutions.

None of these problems is present in the theory given below.

An alternative approach, reviewed here and expanded upon, begins with
the basis vectors carrying the unitary, irreducible representations
of the Poincaré group (the group of spacetime translations and Lorentz
transformations). These basis vectors are constructed to have only
positive energies, in accordance with physical observations. In fact
the operator representing time reversal is antiunitary precisely so
that the positivity of energy is preserved under this transformation.
Normalized superpositions of these basis vectors can be formed to
describe wavepacket states. The relativistic transformations of the
superposition amplitudes are to be derived rather than postulated.
We come to realize that relativity allows these, generally more complicated,
transformation properties (we call this non-manifest covariance) as
well as the manifest covariance of scalars, four-vectors and tensors.

From these complex superposition coefficients we can identify relativistic
momentum/spin-$z$-component probability amplitudes for a massive
particle of any spin. Then we refer to the construction by Newton
and Wigner \cite{Newton1949} of position/spin-$z$-component eigenvectors
for a massive particle with any spin. This will lead us to identify
relativistic position/spin-$z$-component probability amplitudes.
We derive their transformation properties, which are found to be,
for boosts, nonlocal.

In Section II we derive the relativistic transformation properties
of the momentum/spin-$z$-component probability amplitudes for a massive
particle with general spin. Many of these results have already been
explained by Fong and Rowe \cite{Fong1968}. However it appears that
it is still not widely known that probability amplitudes taking the
same roles of those in the nonrelativistic theory can be consistent
with special relativity. Hence we desire to bring their work to more
widespread attention and to expand on it.

In Section III we construct eigenvectors of position and the position
operator, following the work of Newton and Wigner \cite{Newton1949}.
The apparent difference between their position operator and the nonrelativistic
case can be easily understood as being due to the fact that these
two operators are constructed to act on different wavefunctions. They
are, in fact, the same operator when compared using the same wavefunctions.
This fact was also explained by Fong and Rowe \cite{Fong1968}.

For reference, we show how to construct positive energy solutions
of the Klein-Gordon and Dirac equations in terms of probability amplitudes.

We consider the spinless case and find the Lorentz transformation
properties of the position operator and derive some physically meaningful
conclusions about the relativity of position measurements.

In Section IV we discuss the concept of causality in quantum mechanics,
limited, as it is, by the uncertainty principle. In Section V we comment
briefly on the relevance of our results to theories of interaction.
Conclusions follow in Section VI.

Throughout this paper we use Heaviside-Lorentz units, in which $\hbar=c=\epsilon_{0}=\mu_{0}=1$.
We use the active convention for Poincaré transformations, in which,
for example, the boost by velocity $\boldsymbol{\beta}_{0}$ of a
particle of mass $m_{0}$ at rest produces a particle with momentum
$p^{\mu}=m_{0}(\gamma_{0},\gamma_{0}\boldsymbol{\beta}_{0})^{\mu},$
with $\gamma_{0}=1/\sqrt{1-\boldsymbol{\beta}_{0}^{2}}.$

\section{Relativistic momentum-spin probability amplitudes for massive particles}

To calculate the probability of a process in quantum mechanics, a
number between 0 and 1, requires representing the physical states
of the system by state vectors that have a finite normalization. The
expressions for these probabilities take their simplest forms if the
state vectors are normalized to unity, so we follow that convention
throughout. We are interested in cases where one or more of the observables
have eigenvalues with a continuous spectrum. Then the eigenvectors
of such an observable are improper state vectors with delta function
normalizations. To form a state vector with unit normalization requires
constructing a coherent superposition of these basis vectors. From
the complex number coefficients in this superposition can be identified
the relativistic probability amplitudes. These amplitudes, the physical
information they contain, the operators that act on them and the relativistic
transformations of these quantities are the main topics dealt with
here.

The concept of a fundamental particle is intimately linked to that
of irreducible representations of the Poincaré group \cite{Wigner1939,Halpern1968,Newton1949}.
While we believe the electron (and positron) to have no substructure,
to the best of our experimental knowledge, we take as their free momentum/spin-component
basis vectors $|\,p,\frac{1}{2},m\,\rangle,$ labelled by the four-momentum,
$p^{\mu}=(\omega,\boldsymbol{p})^{\mu},$ the spin, $1/2,$ and the
spin $z$ component in the rest frame. These, as we will see, carry
the irreducible representations of the Poincaré group. We leave off
the charge and any other labels. (While we label the basis vectors
by the four-momenta to make their transformations more transparent,
we know that there are only three independent quantum numbers, the
three components of $\boldsymbol{p}$.) If substructure were to be
discovered, using an irreducible representation would become, at best,
an approximation.

For a general spin, $s=0,\frac{1}{2},1,\dots,$ we have basis vectors
$|\,p,s,m\,\rangle$ with spin $z$ component $m=-s,-(s-1),\dots,s-1,s.$
These basis vectors are improper state vectors with delta function
normalization. We choose to use the invariant normalization,
\begin{equation}
\langle\,p_{1},s,m_{1}\,|\,p_{2},s,m_{2}\,\rangle=\delta_{m_{1}m_{2}}\omega_{1}\delta^{3}(\boldsymbol{p}_{1}-\boldsymbol{p}_{2}),\label{eq:2.1}
\end{equation}
as this makes the following transformations take their simplest forms.
These basis vectors carry irreducible unitary representations of spacetime
translations \cite{Wigner1939,Halpern1968}:
\begin{equation}
U(T(a))\,|\,p,s,m\,\rangle=|\,p,s,m\,\rangle e^{+ip\cdot a},\label{eq:2.2}
\end{equation}
rotations (with double-valued representations for half-integral spins):
\begin{equation}
U(R)\,|\,p,s,m\,\rangle=\sum_{m^{\prime}=-s}^{s}|\,Rp,s,m^{\prime}\,\rangle\mathcal{D}_{m^{\prime}m}^{(s)}(R)\label{eq:2.3}
\end{equation}
and boosts:
\begin{equation}
U(\Lambda)\,|\,p,s,m\,\rangle=\sum_{m^{\prime}=-s}^{s}|\,\Lambda p,s,m^{\prime}\,\rangle\mathcal{W}_{m^{\prime}m}^{(s)}(\Lambda p\leftarrow p),\label{eq:2.4}
\end{equation}
the elements of the Poincaré group. In addition, they represent space
inversion:
\begin{equation}
U(\mathcal{P})\,|\,(\omega,\boldsymbol{p}),s,m\,\rangle=\eta\,|\,(\omega,-\boldsymbol{p}),s,m\,\rangle\label{eq:2.5}
\end{equation}
and time reversal
\begin{equation}
A(\mathcal{T})\,|\,(\omega,\boldsymbol{p}),s,m\,\rangle=|\,(\omega,-\boldsymbol{p}),s,-m\,\rangle(-)^{s-m}.\label{eq:2.6}
\end{equation}
Here $\eta=\pm1$ is called the intrinsic parity of the particle.
Note that the antiunitary time reversal operator, $A(\mathcal{T}),$
when applied to a superposition, involves taking the complex conjugate
of the amplitudes. Also $\mathcal{W}$ is a matrix element of a Wigner
rotation, which can be evaluated by
\begin{equation}
W(\Lambda p\leftarrow p)=\Lambda^{-1}[\Lambda p]\,\Lambda\,\Lambda[p].\label{eq:2.7}
\end{equation}
where
\begin{equation}
\Lambda[p]\equiv\Lambda(\frac{\boldsymbol{p}}{\omega})\label{eq:2.8}
\end{equation}
and $\Lambda(\boldsymbol{\beta})$ is a function of the boost velocity,
$\boldsymbol{\beta}.$ Explicit forms of the Wigner rotations are
given in \cite{Halpern1968}. Two successive, noncollinear, boosts
(from the rest momentum to $p$ and then from $p$ to $\Lambda p$)
produce a boost (from the rest momentum to $\Lambda p$) preceded
by a rotation in the rest frame. This is the physics behind the Thomas
precession \cite{Jackson1975}.

As discussed above, we must form coherent superpositions of the improper
basis vectors to make normalized state vectors. We write the general
case as
\begin{equation}
|\,\psi\,\rangle=\int d^{3}p\sum_{m=-s}^{s}|\,p,s,m\,\rangle\frac{1}{\sqrt{\omega}}\Psi_{m}(p).\label{eq:2.9}
\end{equation}
The factor of $1/\sqrt{\omega}$ is to compensate for the covariant
normalization, making this expression just like the familiar nonrelativistic
case. By this we mean that the state vectors
\begin{equation}
|\,p,s,m;\mathrm{NR}\,\rangle=|\,p,s,m\,\rangle\frac{1}{\sqrt{\omega}}\label{eq:2.10}
\end{equation}
have the familiar orthonormality
\begin{equation}
\langle\,p_{1},s,m_{1};\mathrm{NR}\,|\,p_{2},s,m_{2};\mathrm{NR}\,\rangle=\delta^{3}(\boldsymbol{p}_{1}-\boldsymbol{p}_{2})\label{eq:2.11}
\end{equation}
used in nonrelativistic quantum mechanics. (Again we note that $\Psi_{m}(p)$
is a function of only three independent momentum components.)

We also define, for comparison, amplitudes
\begin{equation}
\Phi_{m}(p)=\sqrt{\omega}\,\Psi_{m}(p).\label{eq:2.12}
\end{equation}

Now we can derive (not postulate) the relativistic transformation
properties of the $\Psi_{m}(p)$ (and thus, of the $\Phi_{m}(p)$).
The technique is to apply the unitary (or antiunitary) transformation
to $|\,\psi\,\rangle$ and thus to the basis vectors, then manipulate
the expression into the form
\begin{equation}
U/A\,|\,\psi\,\rangle=\int d^{3}p\sum_{m=-s}^{s}|\,p,s,m\,\rangle\frac{1}{\sqrt{\omega}}\Psi_{m}^{\prime}(p),\label{eq:2.13}
\end{equation}
then extract the $\Psi_{m}^{\prime}(p)$ by orthonormality.

The transformation rules for the Poincaré transformations are:
\begin{eqnarray}
\mathrm{Spacetime\ translations:}\quad\Psi_{m}^{\prime}(p) & = & \Psi_{m}(p)\,e^{+ip\cdot a},\nonumber \\
\mathrm{Rotations:}\quad\Psi_{m}^{\prime}(p) & = & \sum_{m^{\prime}=-s}^{s}\mathcal{D}_{mm^{\prime}}^{(s)}(R)\,\Psi_{m^{\prime}}(R^{-1}p),\nonumber \\
\mathrm{Boosts}:\quad\Psi_{m}^{\prime}(p) & = & \sqrt{\gamma_{0}(1-\boldsymbol{\beta}_{0}\cdot\boldsymbol{\beta})}\sum_{m^{\prime}=-s}^{s}\mathcal{W}_{mm^{\prime}}^{(s)}(p\leftarrow\Lambda^{-1}p)\Psi_{m^{\prime}}(\Lambda^{-1}p),\label{eq:2.14}
\end{eqnarray}
where $\boldsymbol{\beta}_{0}$ is the boost velocity, $\gamma_{0}=1/\sqrt{1-\boldsymbol{\beta}_{0}^{2}}$
and $\boldsymbol{\beta}=\boldsymbol{p}/\omega$ is the velocity of
the particle. For the inversions, we have
\begin{eqnarray}
\mathrm{Space\ inversion:}\quad\Psi_{m}^{\prime}(\omega,\boldsymbol{p}) & = & \eta\,\Psi_{m}(\omega,-\boldsymbol{p}),\nonumber \\
\mathrm{Time\ reversal:}\quad\Psi_{m}^{\prime}(\omega,\boldsymbol{p}) & = & (-)^{s+m}\Psi_{-m}^{*}(\omega,-\boldsymbol{p}).\label{eq:2.15}
\end{eqnarray}

In these equations we have satisfied the requirements of special relativity.
We have the rules for transforming our physical quantity under Poincaré
transformations, rules which depend only on the translation, rotation
and boost parameters. What we have here are examples of \textit{nonmanifest}
covariance.

These transformations all preserve the modulus squared of scalar products:
\begin{equation}
\left|\int d^{3}p^{\prime}\sum_{m^{\prime}=-s}^{s}\Psi_{m^{\prime}}^{(1)\prime*}(p^{\prime})\Psi_{m^{\prime}}^{(2)\prime}(p^{\prime})\right|^{2}=\left|\int d^{3}p\sum_{m=-s}^{s}\Psi_{m}^{(1)*}(p)\Psi_{m}^{(2)}(p)\right|^{2},\label{eq:2.15.1}
\end{equation}
as can be easily verified from Eqs. (\ref{eq:2.14},\ref{eq:2.15}).
For example, we show the $s=0$ boost case. With $p^{\prime}=\Lambda p,$
we have
\begin{equation}
\Psi^{(n)\prime}(p^{\prime})=\sqrt{\gamma_{0}(1-\boldsymbol{\beta}_{0}\cdot\boldsymbol{\beta})}\,\Psi^{(n)}(p)=\sqrt{\frac{\omega}{\omega^{\prime}}}\,\Psi^{(n)}(p),\quad\mathrm{for}\ n=1,2,\label{eq:2.15.2}
\end{equation}
so
\begin{equation}
\int d^{3}p^{\prime}\,\Psi^{(1)\prime}(p^{\prime})\Psi^{(2)\prime}(p^{\prime})=\int\frac{d^{3}p^{\prime}}{\omega^{\prime}}\,\omega\,\Psi^{(1)}(p)\Psi^{(2)}(p)=\int\frac{d^{3}p}{\omega}\,\omega\,\Psi^{(1)}(p)\Psi^{(2)}(p)=\int d^{3}p\,\Psi^{(1)}(p)\Psi^{(2)}(p).\label{eq:2.15.3}
\end{equation}

We point out the important distinction between unitary covariance,
manifest covariance and nonmanifest covariance for Poincaré transformations.
Once the unitary Poincaré transformations of the basis vectors are
specified (such as Eqs. (\ref{eq:2.2},\ref{eq:2.3},\ref{eq:2.4},\ref{eq:2.5},\ref{eq:2.6})),
the unitary transformations of all amplitudes and all observables
written on those basis vectors are specified. Any equation involving
observables, such as $i[A,B]=C,$ will take the same form in all frames,
$i[A^{\prime},B^{\prime}]=C^{\prime},$ with $A^{\prime}=U^{\dagger}(L)AU(L)$
\textit{etc}. A subset of the observables will transform manifestly,
as scalars, four-vectors or tensors. An example is the four-momentum
operator, with $P^{\prime\mu}=U^{\dagger}(L)P^{\mu}U(L)=L_{\phantom{\mu}\nu}^{\mu}P^{\nu}.$
The remainder of the observables will transform nonmanifestly, not
as scalars, four-vectors or tensors. The prime example of this will
be the position operator (Eq. (\ref{eq:3.29}) below). Special relativity
does not demand that all quantities of physical interest transform
with Lorentz indices, merely that all such transformations are well
defined and depend only on the translation, rotation and boost parameters.

The normalization condition becomes, in two forms,
\begin{equation}
1=\int d^{3}p\sum_{m=-s}^{s}|\Psi_{m}(p)|^{2}=\int\frac{d^{3}p}{\omega}\sum_{m=-s}^{s}|\Phi_{m}(p)|^{2}.\label{eq:2.16}
\end{equation}
The expectation of the four-momentum operator, $P^{\mu},$ becomes,
also in two forms
\begin{equation}
\langle\,\psi\,|\,P^{\mu}\,|\,\psi\,\rangle=\int d^{3}p\,p^{\mu}\sum_{m=-s}^{s}|\Psi_{m}(p)|^{2}=\int\frac{d^{3}p}{\omega}\,p^{\mu}\sum_{m=-s}^{s}|\Phi_{m}(p)|^{2}.\label{eq:2.17}
\end{equation}
The first forms in both cases show the probability interpretation
of our probability amplitudes, in exactly the same form as in the
nonrelativistic theory. The quantity
\begin{equation}
\rho(\boldsymbol{p})=\sum_{m=-s}^{s}|\Psi_{m}(p)|^{2}\label{eq:2.18}
\end{equation}
acts like a normalized momentum probability density, with $\rho(\boldsymbol{p})\Delta V(\boldsymbol{p})$
being the probability of measuring the momentum of the particle in
a small volume $\Delta V(\boldsymbol{p})$ of momentum space. Furthermore,
the use of a projector onto an eigenvector of momentum and spin component
in
\begin{equation}
\langle\,\psi\,|\,p,s,m\,\rangle\langle\,p,s,m\,|\,\psi\,\rangle=|\Psi_{m}(p)|^{2}\label{eq:2.19}
\end{equation}
confirms the interpretation of $\Psi_{m}(p)$ as a position/spin-component
probability amplitude.

The second forms in Eqs. (\ref{eq:2.16},\ref{eq:2.17}) show manifestly
the \textit{invariance} of the normalization result and the \textit{covariance}
of the momentum expectation, once we recognize, as is easily seen
from Eqs. (\ref{eq:2.14}) that the quantity
\begin{equation}
S(p)=\sum_{m=-s}^{s}|\Phi_{m}(p)|^{2}\label{eq:2.20}
\end{equation}
transforms as a scalar function under general Lorentz transformations,
$L$:
\begin{equation}
S^{\prime}(p)=S(L^{-1}p).\label{eq:2.21}
\end{equation}

We are not saying that quantities like $\Phi_{m}(p)$ and the Dirac
amplitudes (discussed below) might not be \textit{useful} in constructing
theories of interaction. We are pointing out that they are not probability
amplitudes, while probability amplitudes $\Psi_{m}(p)$, with the
same probability interpretation as in the nonrelativistic theory,
can be formed and have well-defined relativistic transformation properties.

We will not investigate the relativity of spin measurements in this
paper.

\section{The position operator and its relativistic transformations}

\subsection{Position/spin-component probability amplitudes and the position operator}

Newton and Wigner \cite{Newton1949} proposed four conditions that
should be satisfied by a state vector representing a massive particle,
with spin, localized at the origin at time $t=0$. The state vector
is written as a superposition of the basis vectors $|\,p,s,m\,\rangle.$
These conditions are
\begin{enumerate}
\item A spatial translation should produce a vector orthogonal to the original.
\item A linear superposition of localized state vectors produces another
localized state vector.
\item A rotation should be represented by the linear combination of state
vectors from a set: the set should carry a finite-dimensional irreducible
representation of rotations. Also, the actions of space inversion
and time reversal should also produce a state vector in this set.
\item The superposition coefficients should be everywhere regular functions
of the momentum $\boldsymbol{p},$ which can be enforced by requiring
finiteness upon the application of an infinitesimal boost, which involves
a derivative with respect to momentum.
\end{enumerate}
They found
\begin{equation}
|\,\boldsymbol{0},s,m\,\rangle=\int\frac{d^{3}p}{\omega}|\,p,s,m\,\rangle\frac{\sqrt{\omega}}{(2\pi)^{\frac{3}{2}}}.\label{eq:3.1}
\end{equation}
This spatially translates to
\begin{equation}
|\,\boldsymbol{x},s,m\,\rangle=\int\frac{d^{3}p}{\sqrt{\omega}}|\,p,s,m\,\rangle\frac{e^{-i\boldsymbol{p}\cdot\boldsymbol{x}}}{(2\pi)^{\frac{3}{2}}},\label{eq:3.2}
\end{equation}
and satisfies the first condition with the delta function normalization
\begin{equation}
\langle\,\boldsymbol{x}_{1},s,m_{1}\,|\,\boldsymbol{x}_{2},s,m_{2}\,\rangle=\delta_{m_{1}m_{2}}\delta^{3}(\boldsymbol{x}_{1}-\boldsymbol{x}_{2}).\label{eq:3.3}
\end{equation}
Conditions 2 and 3 are clearly satisfied, with
\begin{align}
U(R)\,|\,\boldsymbol{0},s,m\,\rangle & =\sum_{m^{\prime}=-s}^{s}|\,\boldsymbol{0},s,m^{\prime}\,\rangle\mathcal{D}_{m^{\prime}m}^{(s)}(R),\nonumber \\
U(\mathcal{P})\,|\,\boldsymbol{0},s,m\,\rangle & =\eta\,|\,\boldsymbol{0},s,m\,\rangle,\nonumber \\
A(\mathcal{T})\,|\,\boldsymbol{0},s,m\,\rangle & =\sum_{m^{\prime}=-s}^{s}|\,\boldsymbol{0},s,-m\,\rangle(-)^{s-m}.\label{eq:3.4}
\end{align}
 It can be easily verified that condition 4 is satisfied.

Note that we use the same label, $m,$ for the spin component in each
rest frame and the spin component at position $\boldsymbol{x}.$ Clearly
spin measurements commute with position measurements and momentum
measurements.

From Eq. (\ref{eq:3.2}) we see that we have exactly the same position
eigenvectors as in the nonrelativistic theory, since again the factor
of $1/\sqrt{\omega}$ compensates for the covariant normalization.
Then we can define amplitudes that we expect to have an interpretation
as position/spin-component probability amplitudes
\begin{equation}
\psi_{m}(\boldsymbol{x})=\langle\,\boldsymbol{x},s,m\,|\,\psi\,\rangle=\int\frac{d^{3}p}{(2\pi)^{\frac{3}{2}}}\Psi_{m}(p)\,e^{+i\boldsymbol{p}\cdot\boldsymbol{x}}.\label{eq:3.6}
\end{equation}
In the Schrödinger picture we have time-dependent amplitudes
\begin{equation}
\psi_{m}(t,\boldsymbol{x})=\langle\,\boldsymbol{x},s,m\,|\,e^{-iHt}\,|\,\psi\,\rangle=\int\frac{d^{3}p}{(2\pi)^{\frac{3}{2}}}\Psi_{m}(p)\,e^{+i(\boldsymbol{p}\cdot\boldsymbol{x}-\omega t)}.\label{eq:3.7}
\end{equation}

The time evolution of this wavefunction is governed by the (free)
relativistic Schrödinger equation
\begin{equation}
i\frac{\partial}{\partial t}\psi_{m}(t,\boldsymbol{x})=\hat{H}\,\psi_{m}(t,\boldsymbol{x}).\label{eq:3.7.1}
\end{equation}
The Hamiltonian is completely defined in momentum space, where it
has only positive eigenvalues,
\begin{equation}
\hat{H}\,|\,p,s,m\,\rangle=\omega\,|\,p,s,m\,\rangle=+\sqrt{\boldsymbol{p}^{2}+m_{0}^{2}}\,|\,p,s,m\,\rangle.\label{eq:3.7.2}
\end{equation}
It does not have a representation as a differential operator in position
space except in the nonrelativistic limit, where it reduces to
\begin{equation}
\hat{H}\rightarrow m_{0}-\frac{1}{2m_{0}}\nabla^{2}.\label{eq:3.7.3}
\end{equation}
It is perhaps easiest to understand in position space in terms of
Fourier transforms:
\begin{equation}
\hat{H}=\mathcal{\hat{F}}^{*}(x\leftarrow p)\,\sqrt{\boldsymbol{p}^{2}+m_{0}^{2}}\,\hat{\mathcal{F}}(p\leftarrow x),\label{eq:3.7.3.1}
\end{equation}
where
\begin{align}
\hat{\mathcal{F}}(p\leftarrow x)\,f(x) & =\int\frac{d^{3}x}{(2\pi)^{\frac{3}{2}}}e^{ip\cdot x}\,f(x),\nonumber \\
\mathcal{\hat{F}}^{*}(x\leftarrow p)\,F(p) & =\int\frac{d^{3}p}{(2\pi)^{\frac{3}{2}}}e^{-ip\cdot x}\,F(p).\label{eq:3.7.3.2}
\end{align}

We confirm the covariance of the relativistic Schrödinger equation
under boosts. With
\begin{equation}
\psi_{m^{\prime}}^{\prime}(x^{\prime})=\langle\,\boldsymbol{x}^{\prime},m^{\prime}\,|\,e^{-iHt^{\prime}}\,U(\Lambda)\,|\,\psi\,\rangle=\int\frac{d^{3}p^{\prime}}{(2\pi)^{\frac{3}{2}}}\Psi_{m^{\prime}}^{\prime}(p^{\prime})\,e^{-ip^{\prime}\cdot x^{\prime}},\label{eq:3.7.3.3}
\end{equation}
we have
\begin{align}
i\frac{\partial}{\partial t^{\prime}}\psi_{m^{\prime}}^{\prime}(x^{\prime}) & =\int\frac{d^{3}p^{\prime}}{(2\pi)^{\frac{3}{2}}}\,\omega^{\prime}\,\Psi_{m^{\prime}}^{\prime}(p^{\prime})\,e^{-ip^{\prime}\cdot x^{\prime}}\nonumber \\
 & =\hat{\mathcal{F}}^{*}(x^{\prime}\leftarrow p^{\prime})\,\omega^{\prime}\,\hat{\mathcal{F}}(p^{\prime}\leftarrow x^{\prime})\,\psi_{m^{\prime}}^{\prime}(x^{\prime}).\label{eq:3.7.3.4}
\end{align}
So
\begin{equation}
i\frac{\partial}{\partial t^{\prime}}\psi_{m^{\prime}}^{\prime}(t^{\prime},\boldsymbol{x}^{\prime})=\hat{H}^{\prime}\,\psi_{m^{\prime}}^{\prime}(t^{\prime},\boldsymbol{x}^{\prime}),\label{eq:3.7.4.1}
\end{equation}
with
\begin{equation}
\hat{H}^{\prime}=\hat{\mathcal{F}}^{*}(x^{\prime}\leftarrow p^{\prime})\,\sqrt{\boldsymbol{p}^{\prime2}+m_{0}^{2}}\,\hat{\mathcal{F}}(p^{\prime}\leftarrow x^{\prime}).\label{eq:3.7.4.2}
\end{equation}

The relativistic Schrödinger equation, Eq. (\ref{eq:3.7.1}), can
be seen as the zero component of the equation
\begin{equation}
i\partial^{\mu}\psi_{m}(x)=\hat{P}^{\mu}\psi_{m}(x),\label{eq:3.7.4.3}
\end{equation}
with
\begin{equation}
\hat{\boldsymbol{P}}=\mathcal{\hat{F}}^{*}(x\leftarrow p)\,\boldsymbol{p}\,\hat{\mathcal{F}}(p\leftarrow x).\label{eq:3.7.4.4}
\end{equation}
We note that the position probability amplitude also satisfies the
Klein-Gordon equation (\ref{eq:1.1}).

Since Eq. (\ref{eq:3.7}) is a Fourier transform, the Parseval theorem
tells us that the normalization condition can be written
\begin{equation}
\int d^{3}x\sum_{m=-s}^{s}|\psi_{m}(t,x)|^{2}=\int d^{3}p\sum_{m=-s}^{s}|\Psi_{m}(p)|^{2}=1,\label{eq:3.8}
\end{equation}
which is a Poincaré (and inversion) invariant statement, as we have
already seen that fact for the momentum integral.

Then we can find the Poincaré transformation (and inversion) properties
of these amplitudes:
\begin{eqnarray}
\mathrm{Spacetime\ Translations}:\quad\psi_{m}^{\prime}(x) & = & \psi_{m}(x-a),\nonumber \\
\mathrm{Rotations:}\quad\psi_{m}^{\prime}(x) & = & \sum_{m^{\prime}=-s}^{s}\mathcal{D}_{mm^{\prime}}^{(s)}(R)\psi_{m^{\prime}}(R^{-1}x),\nonumber \\
\mathrm{Boosts:}\quad\psi_{m}^{\prime}(x) & = & \sum_{m^{\prime}=-s}^{s}\hat{B}_{mm^{\prime}}\psi_{m^{\prime}}(\Lambda^{-1}x),\nonumber \\
\mathrm{Space\ inversion:}\quad\psi_{m}^{\prime}(t,\boldsymbol{x}) & = & \eta\,\psi_{m}(t,-\boldsymbol{x}),\nonumber \\
\mathrm{Time\ reversal:}\quad\psi_{m}^{\prime}(t,\boldsymbol{x}) & = & (-)^{s+m}\,\psi_{-m}^{*}(-t,\boldsymbol{x}).\label{eq:3.9}
\end{eqnarray}
The operator appearing in the boost transformation is, in terms of
Fourier transforms,
\begin{equation}
\hat{B}_{mm^{\prime}}=\hat{\mathcal{F}}^{*}(x\leftarrow p)\,\sqrt{\gamma_{0}(1-\boldsymbol{\beta}_{0}\cdot\boldsymbol{\beta})}\,\mathcal{W}_{mm^{\prime}}^{(s)}(p\leftarrow\Lambda^{-1}p)\,\hat{\mathcal{F}}(\Lambda^{-1}p\leftarrow\Lambda^{-1}x).\label{eq:3.9.1}
\end{equation}
This transformation is nonlocal, as it must be to preserve probability.

These transformations all preserve the modulus squared of scalar products:
\[
\left|\int d^{3}x^{\prime}\sum_{m^{\prime}=-s}^{s}\psi_{m}^{(1)\prime*}(x^{\prime})\psi_{m}^{(2)\prime}(x^{\prime})\right|^{2}=\left|\int d^{3}x\sum_{m=-s}^{s}\psi_{m}^{(1)*}(x)\psi_{m}^{(2)}(x)\right|^{2},
\]
as can be easily verified from Eqs. (\ref{eq:3.9}).

Since at $t=0$ the Fourier transform inverse to Eq. (\ref{eq:3.6})
is
\begin{equation}
\Psi_{m}(p)=\int\frac{d^{2}x}{(2\pi)^{\frac{3}{2}}}\psi_{m}(0,\boldsymbol{x})\,e^{-i\boldsymbol{p}\cdot\boldsymbol{x}},\label{eq:3.10}
\end{equation}
the operator in momentum space that produces a factor of $\boldsymbol{x},$
the position operator at $t=0,$ is simply
\begin{equation}
\hat{\boldsymbol{x}}=i\frac{\partial}{\partial\boldsymbol{p}},\label{eq:3.11}
\end{equation}
as in the nonrelativistic theory, where $\partial/\partial\boldsymbol{p}\equiv(\partial/\partial p_{x},\partial/\partial p_{y},\partial/\partial p_{z}).$
Since $i[\hat{H},\hat{\boldsymbol{x}}]=\hat{\boldsymbol{\beta}}$
and $i[\hat{H},\hat{\boldsymbol{\beta}}]=0,$ we have, in the Heisenberg
picture,
\begin{equation}
\hat{\boldsymbol{x}}(t)=e^{+iHt}\,\hat{\boldsymbol{x}}\,e^{-iHt}=\hat{\boldsymbol{x}}+\hat{\boldsymbol{\beta}}t,\label{eq:3.11.1}
\end{equation}
where $\hat{\boldsymbol{\beta}}$ is the operator with eigenvalue
$\boldsymbol{p}/\omega$ acting on a momentum/spin-component eigenvector.

Newton and Wigner \cite{Newton1949} find the position operator to
be
\begin{equation}
\hat{\boldsymbol{x}}_{\mathrm{NW}}=i\frac{\partial}{\partial\boldsymbol{p}}-i\frac{\boldsymbol{p}}{2\omega^{2}},\label{eq:3.11.2}
\end{equation}
but theirs is constructed to act on $\Phi_{m}(p)$ wavefunctions rather
than $\Psi_{m}(p)$ probability amplitudes, with the invariant measure
in their scalar products. The apparent difference is resolved in the
identity
\begin{equation}
\int\frac{d^{3}p}{\omega}\sum_{m=-s}^{s}\Phi_{m}^{(1)*}(p)\{i\frac{\partial}{\partial\boldsymbol{p}}-i\frac{\boldsymbol{p}}{2\omega^{2}}\}\Phi_{m}^{(2)}(p)=\int d^{3}p\sum_{m=-s}^{s}\Psi_{m}^{(1)*}(p)\{i\frac{\partial}{\partial\boldsymbol{p}}\}\Psi_{m}^{(2)}(p).\label{eq:3.12}
\end{equation}
The Hermiticity of the position operator is most easily confirmed
in the second form.

The use of a projector onto an eigenvector of position and spin component
in (Schrödinger picture)
\begin{equation}
\langle\,\psi(t)\,|\,\boldsymbol{x},m\,\rangle\langle\,\boldsymbol{x},m\,|\,\psi(t)\,\rangle=|\psi_{m}(t,x)|^{2}\label{eq:3.13}
\end{equation}
confirms the interpretation of $\psi_{m}(x)$ as a position/spin-component
probability amplitude.

Note that we deal only with positive energy solutions of the relativistic
Schrödinger equation, Eq. (\ref{eq:3.7.1}),and that the action of
the position operator does not introduce negative energies. This was
discussed by Newton and Wigner \cite{Newton1949}, who pointed out
that the argument that measuring the position of an electron in a
very small volume would lead to pair creation is not consistent with
this construction of the position operator. While pair creation would
occur if high energy photons were used to localize the electron, the
simplest way to localize an electron in a very small volume is to
observe it from a frame boosted by a highly relativistic velocity
relative to the average rest frame. Lorentz contraction will localize
the electron in an arbitrarily small volume without pair creation,
as can be verified in the rest frame. Lorentz contraction and the
spreading with time of relativistic wavepackets will be dealt with
in a paper in preparation.

We also note that our results for spin other than zero are different
from those of Newton and Wigner \cite{Newton1949}, since they derived
the action of their position operator on Dirac amplitudes.

\subsection{Locally transforming position amplitudes}

From the probability amplitude, $\Psi(p),$ for a massive, spinless
particle, we can construct an amplitude that transforms locally as
a scalar function of $x^{\mu}=(t,\boldsymbol{x})^{\mu},$
\[
\varphi(x)=\int\frac{d^{3}p}{\omega}e^{-ip\cdot x}\,\sqrt{\omega}\,\Psi(p),
\]
and is a positive energy solution of the Klein-Gordon equation.

From the probability amplitudes, $\Psi_{\pm\frac{1}{2}}(p),$ for
an electron, we can construct a positive energy solution of the Dirac
equation,
\[
\sum_{b=1}^{4}[i\gamma^{\mu}\partial_{\mu}-m_{e}]_{ab}\Psi_{b}^{\mathrm{Dirac}}(x)=0\quad\mathrm{for}\ a=1,\dots,4,
\]
with the gamma matrices ($\gamma^{\mu}$) in the Weyl or chiral representation
\cite{Itzykson1980}. This is the four-component column vector
\[
\Psi^{\mathrm{Dirac}}(x)=\int\frac{d^{3}p}{\omega}e^{-ip\cdot x}\,\frac{1}{\sqrt{2}}\begin{pmatrix}\mathrm{D}_{+\frac{1}{2},m}^{(-)}[p]\\
\mathrm{D}_{-\frac{1}{2},m}^{(-)}[p]\\
\mathrm{D}_{+\frac{1}{2},m}^{(+)}[p]\\
\mathrm{D}_{-\frac{1}{2},m}^{(+)}[p]
\end{pmatrix}\sqrt{\omega}\,\Psi_{m}(p),
\]
where
\[
\mathrm{D}^{(r)}[p]=\sqrt{\frac{\omega+m_{e}}{2m_{e}}}+r\sqrt{\frac{\omega-m_{e}}{2m_{e}}}\hat{\boldsymbol{p}}\cdot\boldsymbol{\sigma}\quad\mathrm{for}\,r=\pm1
\]
and $\boldsymbol{\sigma}$ are the $2\times2$ Pauli matrices. These
components transform locally under a four-dimensional nonunitary transformation.
Weinberg \cite{Weinberg1964} shows how to construct locally transforming
amplitudes for any spin, but his interest was not in probability amplitudes.

\subsection{The boost transformation of the position operator}

Note that the momentum dependence of the Wigner rotation (the boost
case of Eq. (\ref{eq:2.14})) means that the position wavefunction
of a particle with spin will change under boosts in a way different
from the spinless case. We restrict our attention here to the spinless
case and derive the boost transformation of the position operator.

In what follows we neglect to put hats on operators, using them instead
for unit vectors. The unitary boost transformation can be written
\begin{equation}
U(\Lambda)=e^{-i\boldsymbol{\zeta}\cdot\boldsymbol{K}},\label{eq:3.14}
\end{equation}
where $\boldsymbol{\zeta}$ is called the rapidity, satisfying
\begin{equation}
\cosh\zeta=\gamma_{0},\quad\sinh\zeta\,\hat{\boldsymbol{\zeta}}=\gamma_{0}\boldsymbol{\beta}_{0},\label{eq:3.15}
\end{equation}
and the boost generator is
\begin{equation}
\boldsymbol{K}=-\frac{i}{2}\omega\frac{\partial}{\partial\boldsymbol{p}}-\frac{i}{2}\frac{\partial}{\partial\boldsymbol{p}}\omega=-\frac{1}{2}\{\omega,\boldsymbol{x}\},\label{eq:3.16}
\end{equation}
found by considering Eq. (\ref{eq:2.14}) for an infinitesimal transformation.
Here $\boldsymbol{K}$ is the operator defined to act on the $\Psi(p)$
wavefunctions. (The anticommutator of two Hermitian operators is defined
as $\{A,B\}=AB+BA$ and is clearly Hermitian.)

We start by separating the position operator into parts parallel and
perpendicular to the boost direction:
\begin{equation}
\boldsymbol{x}=\boldsymbol{x}_{\parallel}+\boldsymbol{x}_{\perp},\label{eq:3.17}
\end{equation}
with
\begin{equation}
\boldsymbol{x}_{\parallel}=\boldsymbol{x}\cdot\hat{\boldsymbol{\zeta}}\hat{\boldsymbol{\zeta}}.\label{eq:3.18}
\end{equation}
We note that
\begin{equation}
\hat{\boldsymbol{\zeta}}\boldsymbol{\zeta}\cdot\boldsymbol{K}=-\frac{\zeta}{2}\{\omega,\boldsymbol{x}_{\parallel}\},\label{eq:3.19}
\end{equation}
and this operator will clearly be invariant under this Lorentz transformation.
This gives
\begin{equation}
\{\omega^{\prime},\boldsymbol{x}_{\parallel}^{\prime}\}=\{\omega,\boldsymbol{x}_{\parallel}\}.\label{eq:3.20}
\end{equation}
This equation is easily solved to give
\begin{equation}
\boldsymbol{x}_{\parallel}^{\prime}=\frac{1}{2}\left\{ \frac{1}{\gamma_{0}(1+\boldsymbol{\beta}_{0}\cdot\boldsymbol{\beta})},\boldsymbol{x}_{\parallel}\right\} ,\label{eq:3.21}
\end{equation}
in manifestly Hermitian form.

For the perpendicular components, we define two unit vectors so that
$\{\hat{\boldsymbol{u}}_{1},\hat{\boldsymbol{u}}_{2},\hat{\boldsymbol{\zeta}}\}$
is a right-handed set of three mutually perpendicular axes. Then we
calculate
\begin{equation}
\hat{\boldsymbol{u}}_{1}\cdot\boldsymbol{K}^{\prime}=e^{+i\boldsymbol{\zeta}\cdot\boldsymbol{K}}\,\hat{\boldsymbol{u}}_{1}\cdot\boldsymbol{K}\,e^{-i\boldsymbol{\zeta}\cdot\boldsymbol{K}}=\hat{\boldsymbol{u}}_{1}\cdot\boldsymbol{K}+i\zeta[\hat{\boldsymbol{\zeta}}\cdot\boldsymbol{K},\hat{\boldsymbol{u}}_{1}\cdot\boldsymbol{K}]-\frac{1}{2}\zeta^{2}[\hat{\boldsymbol{\zeta}}\cdot\boldsymbol{K},[\hat{\boldsymbol{\zeta}}\cdot\boldsymbol{K},\hat{\boldsymbol{u}}_{1}\cdot\boldsymbol{K}]]+\dots\label{eq:3.22}
\end{equation}
by using the Lorentz group commutators \cite{Jackson1975}
\begin{equation}
[K_{i},K_{j}]=-i\epsilon_{ijk}J_{k},\quad[K_{i},J_{j}]=+i\epsilon_{ijk}K_{k}.\label{eq:3.23}
\end{equation}
A similar calculation gives $\hat{\boldsymbol{u}}_{2}\cdot\boldsymbol{K}^{\prime}.$
We find
\begin{align}
\hat{\boldsymbol{u}}_{1}\cdot\boldsymbol{K}^{\prime} & =\gamma_{0}\,\hat{\boldsymbol{u}}_{1}\cdot\boldsymbol{K}+\gamma_{0}\beta_{0}\,\hat{\boldsymbol{u}}_{2}\cdot\boldsymbol{J},\nonumber \\
\hat{\boldsymbol{u}}_{2}\cdot\boldsymbol{K}^{\prime} & =\gamma_{0}\,\hat{\boldsymbol{u}}_{2}\cdot\boldsymbol{K}-\gamma_{0}\beta_{0}\,\hat{\boldsymbol{u}}_{1}\cdot\boldsymbol{J}.\label{eq:3.24}
\end{align}
Using $\boldsymbol{J}=\boldsymbol{x}\times\boldsymbol{p},$ we find
\begin{align}
\hat{\boldsymbol{u}}_{2}\cdot\boldsymbol{J} & =\hat{\boldsymbol{u}}_{1}\cdot\boldsymbol{p}\,\hat{\boldsymbol{\zeta}}\cdot\boldsymbol{x}-p_{\parallel}\,\hat{\boldsymbol{u}}_{1}\cdot\boldsymbol{x},\nonumber \\
-\hat{\boldsymbol{u}}_{1}\cdot\boldsymbol{J} & =\hat{\boldsymbol{u}}_{2}\cdot\boldsymbol{p}\,\hat{\boldsymbol{\zeta}}\cdot\boldsymbol{x}-p_{\parallel}\,\hat{\boldsymbol{u}}_{2}\cdot\boldsymbol{x}.\label{eq:3.25}
\end{align}
Together, these results give
\begin{equation}
\boldsymbol{K}_{\perp}^{\prime}=\gamma_{0}\boldsymbol{K}_{\perp}+\gamma_{0}(\boldsymbol{p}_{\perp}\,\boldsymbol{\beta}_{0}\cdot\boldsymbol{x}-\boldsymbol{\beta}_{0}\cdot\boldsymbol{p}\,\boldsymbol{x}_{\perp}).\label{eq:3.26}
\end{equation}

Solving this equation for $\boldsymbol{x}_{\perp}^{\prime},$ using
\begin{equation}
\boldsymbol{K}_{\perp}^{\prime}=-\frac{1}{2}\{\omega,\boldsymbol{x}_{\perp}^{\prime}\},\label{eq:3.27}
\end{equation}
\textit{etc}. gives
\begin{equation}
\boldsymbol{x}_{\perp}^{\prime}=\boldsymbol{x}_{\perp}-\frac{1}{2}\left\{ \frac{1}{1+\boldsymbol{\beta}_{0}\cdot\boldsymbol{\beta}},\boldsymbol{\beta}_{\perp}\,\boldsymbol{\beta}_{0}\cdot\boldsymbol{x}\right\} .\label{eq:3.28}
\end{equation}

So the transformation law is
\begin{equation}
\boldsymbol{x}^{\prime}=\boldsymbol{x}_{\perp}-\frac{1}{2}\left\{ \frac{1}{1+\boldsymbol{\beta}_{0}\cdot\boldsymbol{\beta}},\boldsymbol{\beta}_{\perp}\,\boldsymbol{\beta}_{0}\cdot\boldsymbol{x}\right\} +\frac{1}{2}\left\{ \frac{1}{\gamma_{0}(1+\boldsymbol{\beta}_{0}\cdot\boldsymbol{\beta})},\boldsymbol{x}_{\parallel}\right\} .\label{eq:3.29}
\end{equation}

We know that the definition of velocity, $\boldsymbol{\beta}=i[H,\boldsymbol{x}],$
must take the same form in all frames. However, it was of value to
verify this result using the explicit form of the transformation of
the position operator, as a check on Eq.~(\ref{eq:3.29}). We found
\begin{equation}
i[H^{\prime},\boldsymbol{x}^{\prime}]=i[\gamma_{0}(H+\boldsymbol{\beta}_{0}\cdot\boldsymbol{P}),\boldsymbol{x}^{\prime}]=\frac{\boldsymbol{\beta}_{\perp}+\gamma_{0}(\boldsymbol{\beta}_{\parallel}+\boldsymbol{\beta}_{0})}{\gamma_{0}(1+\boldsymbol{\beta}_{0}\cdot\boldsymbol{\beta})}=\boldsymbol{\beta}^{\prime},\label{eq:3.30}
\end{equation}
the law for the relativistic transformation of velocity \cite{Rindler1977}.

Then we verified the \textit{invariance} of the fundamental commutator
between position and momentum
\begin{equation}
[x_{i}^{\prime},P_{j}^{\prime}]=[x_{i},P_{j}]=i\delta_{ij}.\label{eq:3.31}
\end{equation}
Because the position and momentum have indices, we might have thought
that this commutator transformed like components of a tensor.

\subsection{Lorentz transformations of ``average events''}

We examine the concept of an event in this theory. Since there is
no time operator in quantum mechanics, the position operator cannot
transform as the spatial components of a four-vector. We have derived
its transformation formula and it is linear and homogeneous in the
position operator components with nothing like a time operator appearing.
What, though, of expectation values? If we define
\begin{equation}
x^{\mu}=(t,\langle\,\psi\,|\,\boldsymbol{x}(t)\,|\,\psi\,\rangle)^{\mu}\quad\mathrm{and}\quad x^{\prime\mu}=(t^{\prime},\langle\,\psi\,|\,\boldsymbol{x}^{\prime}(t^{\prime})\,|\,\psi\,\rangle)^{\mu}\label{eq:3.32}
\end{equation}
with
\begin{equation}
t^{\prime}=\gamma_{0}(t+\boldsymbol{\beta}_{0}\cdot\langle\,\psi\,|\,\boldsymbol{x}(t)\,|\,\psi\,\rangle)\label{eq:3.33}
\end{equation}
and
\begin{equation}
\boldsymbol{x}^{\prime}(0)=U^{\dagger}(\Lambda)\,\boldsymbol{x}(0)\,U(\Lambda),\label{eq:3.33.1}
\end{equation}
do we find
\begin{equation}
x^{\prime\mu}=\Lambda_{\phantom{\mu}\nu}^{\mu}x^{\nu}?\label{eq:3.34}
\end{equation}
Here, in the Heisenberg picture, we have
\begin{equation}
\boldsymbol{x}(t)=e^{+iHt}\,\boldsymbol{x}\,e^{-iHt}=\boldsymbol{x}+\boldsymbol{\beta}t\quad\mathrm{and}\quad\boldsymbol{x}^{\prime}(t^{\prime})=e^{+iHt^{\prime}}\,\boldsymbol{x}^{\prime}\,e^{-iHt^{\prime}}=\boldsymbol{x}^{\prime}+\boldsymbol{\beta}^{\prime}t^{\prime},\label{eq:3.35}
\end{equation}
with $\boldsymbol{\beta}^{\prime}$ as in Eq. (\ref{eq:3.30}).

We found the result to be correct, but only as a good approximation
for wavepackets that are narrow in momentum space. The proof involves
replacing operators by their expectation values, which is seen to
be valid to a good approximation using wavepackets of this form. A
typical term to be considered is
\begin{equation}
\boldsymbol{T}=\langle\,\psi\,|\,\frac{1}{2}\left\{ \frac{1}{\gamma_{0}(1+\boldsymbol{\beta}_{0}\cdot\boldsymbol{\beta})},\boldsymbol{x}_{\parallel}\right\} \,|\,\psi\,\rangle.\label{eq:3.36}
\end{equation}
After an integration by parts in which the boundary terms are required
to vanish, this takes the form
\begin{equation}
\boldsymbol{T}=\int d^{3}p\,\frac{1}{2}\left\{ \Psi^{*}(p)\,f(\boldsymbol{p})\,\{i\frac{\partial}{\partial\boldsymbol{p}_{\parallel}}\Psi(p)\}-\{i\frac{\partial}{\partial\boldsymbol{p}_{\parallel}}\Psi^{*}(p)\}\,f(\boldsymbol{p})\,\Psi(p)\right\} ,\label{eq:3.37}
\end{equation}
with
\begin{equation}
f(\boldsymbol{p})=\frac{1}{\gamma_{0}(1+\boldsymbol{\beta}_{0}\cdot\boldsymbol{\beta})}.\label{eq:3.38}
\end{equation}

To proceed, we consider Gaussian wavepackets of the form
\begin{equation}
\Psi(p)=\frac{e^{-|\boldsymbol{p}-\bar{\boldsymbol{p}}|^{2}/4\sigma_{p}^{2}}}{(2\pi\sigma_{p}^{2})^{\frac{3}{4}}}e^{-i\boldsymbol{p}\cdot\bar{\boldsymbol{x}}}.\label{eq:3.39}
\end{equation}
A more general result would be desirable, but is not available at
this point. Since
\begin{equation}
i\frac{\partial}{\partial\boldsymbol{p}}\Psi(p)=\{-i\frac{\boldsymbol{p}-\bar{\boldsymbol{p}}}{2\sigma_{p}^{2}}+\bar{\boldsymbol{x}}\}\Psi(p),\label{eq:3.39.1}
\end{equation}
this leads to
\begin{equation}
\boldsymbol{T}=\bar{\boldsymbol{x}}_{\parallel}\,\int d^{3}p\,|\Psi(p)|^{2}f(\boldsymbol{p}).\label{eq:3.40}
\end{equation}
Since the momentum probability density is a narrow function of momentum
peaked at $\boldsymbol{p}=\bar{\boldsymbol{p}},$ while $f(\boldsymbol{p})$
is slowly varying on $|\boldsymbol{p}-\bar{\boldsymbol{p}}|\apprle\sigma_{p},$
we expand the latter in powers of $\boldsymbol{p}-\bar{\boldsymbol{p}}.$
The integral of the first-order term vanishes by the spherical symmetry
of the momentum probability density around $\boldsymbol{p}=\bar{\boldsymbol{p}}.$
Then we find an upper bound on the fractional remainder, $\varepsilon,$
in
\begin{equation}
\boldsymbol{T}=\frac{\bar{\boldsymbol{x}}_{\parallel}}{\gamma_{0}(1+\boldsymbol{\beta}_{0}\cdot\bar{\boldsymbol{\beta}})}\{1+\varepsilon\}\label{eq:3.41}
\end{equation}
(with $\bar{\boldsymbol{\beta}}=\bar{\boldsymbol{p}}/\sqrt{\bar{\boldsymbol{p}}^{2}+m_{0}^{2}}$)
of
\begin{equation}
|\varepsilon|\apprle\bar{\boldsymbol{\beta}}^{2}\left(\frac{\sigma_{p}}{|\bar{\boldsymbol{p}}|}\right)^{2},\label{eq:3.42}
\end{equation}
required to be much less than unity.

It is important to note that these results on two-component relativistic
probability amplitudes for electrons and positrons in no way contradict
Dirac's construction \cite{Itzykson1980} of a Hermitian current \textit{operator}
for electrons and positrons, which transforms locally as a four-vector
function of position and time and has a zero component that is negative
definite (electrons) or positive definite (positrons). We note that
it was necessary in the construction to use locally transforming \textit{four}-component
objects. These objects are not probability amplitudes.

Similarly, the four-component Dirac \textit{amplitudes} are very useful
in solving the hydrogen atom spectrum \cite{Messiah1961}. However,
the physical meaning of the components is unclear, their values change
in different gamma matrix representations and they are clearly not
probability amplitudes.

In fact these results complement each other. To find the expectation
value of the Dirac current (which is written on the positive energy
$|\,p,\frac{1}{2},m\,\rangle$ basis) and to be able to interpret
the results requires use of the relativistic probability amplitudes
to form normalized state vectors. There is no need to decide whether
position measurements or charge density measurements for an electron
are the ``right'' measurements relativistically. They must be seen
as complementary measurements, with the construction of the detectors
expected to be quite different in the two cases.

\section{Causality in quantum mechanics}

The concept of causality in relativistic quantum mechanics is limited
by the uncertainty principle and the inherent probabilistic nature
of the quantum world. A particle localized at a point at a particular
time is represented by an improper state vector, not part of the physical
Hilbert space. A particle in a physical state always has a nonvanishing
spatial extent over which it can be detected. Thus it is not possible
to define a light cone with exact boundaries for physical particle
events.

The probability distribution in the velocity, $\boldsymbol{\beta},$
of a massive particle vanishes for $|\boldsymbol{\beta}|\geq1.$ So
individual measurements of speed will always give a result less than
the speed of light, and the expectation value of velocity will always
be less than unity in magnitude. These results are reassuring, but
stronger definitions of causality have been proposed and found to
be violated.

Hegerfeldt's criteria \cite{Hegerfeldt1974,Hegerfeldt1985} for causality
ask that the \textit{probability} \textit{density} propagate at less
than the speed of light. The position probability amplitudes presented
here fail his test. We acknowledge that some readers will consider
this violation to be a flaw in our theory. Example position probability
densities have been presented \cite{Rosenstein1987} that violate
this concept of causality quite severely.

These violating state vectors are typically such that the expectations
of higher powers of $|\boldsymbol{x}|$ or $|\boldsymbol{p}|$ are
infinite, and as such may not be physically realizable. More research
needs to be done on the shapes of wavepackets that can be produced
in nature, for example from radioactive decay. Experimental verification
of a small violation of a classical causality condition would be of
great interest. We argue that this would be received as a feature
of probabilistic quantum mechanics rather than a failure of relativity.

We argue that the Newton-Wigner definition of the position operator
is the only correct choice and that Hegerfeldt's criteria for causality
are too strict. In a probabilistic theory, it should be sufficient
that a classical concept of causality be only violated with small
probabilities.

An example illustrates this point. We consider a Gaussian momentum
wavefunction
\[
\Psi(\boldsymbol{\kappa})=\frac{e^{-\boldsymbol{\kappa}^{2}/4}}{(2\pi)^{\frac{3}{4}}},
\]
normalized to $\int d^{3}\kappa\,|\Psi(\boldsymbol{\kappa})|^{2}=1,$
with the dimensionless scaled momentum $\boldsymbol{\kappa}=\boldsymbol{p}/\sigma_{p}.$
We choose the momentum width $\sigma_{p}\gg m_{0},$ giving a very
small spatial width ($\sigma_{x}=1/2\sigma_{p}\ll1/m_{0}$, much smaller
than the Compton wavelength) at $t=0.$ This approximates the massless
case and the superposition is dominated by highly relativistic momenta.
We find the approximate time-dependent spatial wavefunction, spherically
symmetric \cite{Gradsteyn1980} (3.462.1, 9.246 and 9.248.1),
\[
\psi(\tau,\boldsymbol{\rho})=\frac{-i}{(2\pi)^{\frac{3}{4}}}\frac{1}{\sqrt{\pi}}\frac{1}{\rho}\{e^{-(\tau-\rho)^{2}/8}D_{-2}(i(\tau-\rho)/\sqrt{2})-e^{-(\tau+\rho)^{2}/8}D_{-2}(i(\tau+\rho)/\sqrt{2})\}.
\]
normalized to $\int d^{3}\rho\,|\psi(\tau,\boldsymbol{\rho})|^{2}=1,$
with $\boldsymbol{\rho}=\boldsymbol{r}/\sigma_{x}$ and $\tau=t/\sigma_{x}.$
The asymptotic form of this wavefunction as $\rho\rightarrow\infty$
is
\[
\psi(\tau,\boldsymbol{\rho})\rightarrow\frac{1}{(2\pi)^{\frac{3}{4}}}e^{-\rho^{2}/4},
\]
which is stronger localization than was considered in Hegerfeldt's
second theorem \cite{Hegerfeldt1985}. The expectations of all powers
of $|\boldsymbol{\rho}|$ and $|\boldsymbol{\kappa}|$ are finite
for these wavefunctions.

We invoke a causality test very similar to that proposed by Rosenstein
and Usher \cite{Rosenstein1987}. We define the causality ratio, for
$\tau>0,$
\[
C(\tau,\rho)=\frac{\int_{0}^{\rho+\tau}4\pi\rho^{\prime2}d\rho^{\prime}\,|\psi(\tau,\boldsymbol{\rho}^{\prime})|^{2}}{\int_{0}^{\rho}4\pi\rho^{\prime2}d\rho^{\prime}\,|\psi(0,\boldsymbol{\rho}^{\prime})|^{2}}
\]
and require it to be everywhere greater than or equal to unity to
satisfy classical causality, so that no probability leaks out of the
light cone. Numerical results are shown in Figure 1.

\begin{figure}
\noindent \begin{centering}
\includegraphics[width=12cm]{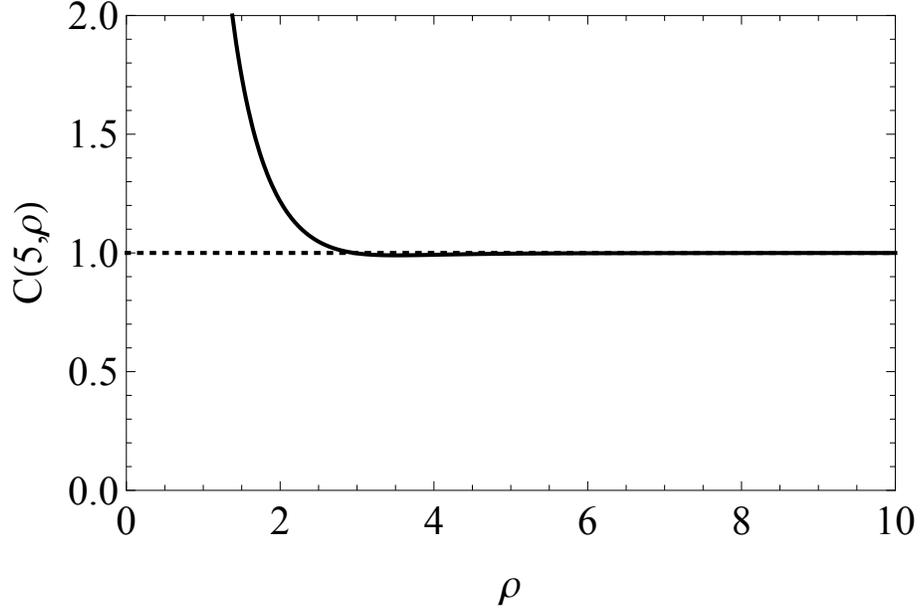}
\par\end{centering}
\caption{The causality ratio $C(5,\rho).$}
\end{figure}

While we would say that the criterion appears to be well satisfied,
it is relevant to note that there is actually a region of tiny violations.
We find $C(5,5)=0.996958.$ So even this particularly well-behaved
wavepacket fails the strict application of a causality criterion,
but would certainly pass a looser, probabilistic interpretation that
we argue is required of quantum mechanics.

\section{Interaction}

Of course everything presented above is for \textit{free} particles,
or particles isolated far from other particles with which they might
interact. It is not our intention in this paper to discuss how to
construct a relativistic theory of interactions involving these probability
amplitudes. Many would say that we already have a relativistic theory
of interaction in relativistic quantum field theory (QFT). We argue,
though, that probability amplitudes are a necessary \textit{addition}
to quantum field theory. That theory calculates elements of the $S$
matrix between plane wave states (momentum eigenvectors), of the example
form
\begin{equation}
\mathcal{M}=\langle\,p_{3},m_{3};p_{4},m_{4}\,|\,S\,|\,p_{1},m_{1};p_{2},m_{2}\,\rangle.\label{eq:6.1}
\end{equation}
To calculate a probability, between 0 and 1, requires construction
of normalized state vectors, which must involve relativistic probability
amplitudes. Thus an example probability would be
\begin{equation}
P=\int\frac{d^{3}p_{1}}{\sqrt{\omega_{1}}}\,\Psi_{m_{1}}(p_{1})\int\frac{d^{3}p_{2}}{\sqrt{\omega_{2}}}\,\Psi_{m_{2}}(p_{2})\int\frac{d^{3}p_{3}}{\sqrt{\omega_{3}}}\,\Psi_{m_{3}}^{*}(p_{3})\int\frac{d^{3}p_{4}}{\sqrt{\omega_{4}}}\,\Psi_{m_{4}}^{*}(p_{4})\langle\,p_{3},m_{3};p_{4},m_{4}\,|\,S\,|\,p_{1},m_{1};p_{2},m_{2}\,\rangle.\label{eq:6.2}
\end{equation}

Wavepackets can regulate divergences. This author presented a wavepacket
treatment of Coulomb scattering \cite{Hoffmann2017a}. The use of
wavepackets introduces a convergence factor into the partial wave
series, which otherwise diverges for all scattering angles.

We are not saying that wavepackets alone can tame the infinities of
relativistic quantum field theory. In Eq.~(\ref{eq:6.2}) above,
we assume that the $S$ matrix element has already been renormalized
before applying the wavepacket superposition. The infinities of QFT
are more deep-rooted than that.

We do present an argument for the consideration of the reader. Suppose,
in a purely quantum mechanical treatment, that we have a Hamiltonian
that can be diagonalized, with finite energies. Then the evolution
operator, an exponential of the Hamiltonian, is also diagonal in this
basis, and is unitary with finite matrix elements. Then the central
problem becomes how to construct incoming and outgoing normalized
state vectors to represent particles localized far from each other.
These state vectors would be written as superpositions of the eigenvectors
of the Hamiltonian, using relativistic probability amplitudes. If
this can be done, we can evolve the incoming state vector in time,
then calculate the probability
\[
P=|\langle\,\mathrm{out}\,|\,e^{-iHT}\,|\,\mathrm{in}\,\rangle|^{2}.
\]
From unitarity, the normalization of the state vectors and Schwartz's
inequality, this probability is guaranteed to be finite, and between
0 and 1.

In the opinion of this author, we do not yet have a quantum mechanical,
relativistic and \textit{finite} theory of interaction with these
properties.

\section{Conclusions}

We have defined momentum/spin-component probability amplitudes and
position/spin-component probability amplitudes for a massive particle
of general spin. We have found their transformation properties under
spacetime translations, general Lorentz transformations, space inversion
and time reversal. We have defined the position operator and derived
its relativistic transformation properties. The results are all very
close to what is done in nonrelativistic quantum mechanics. This should
come as no surprise, since any relativistic theory must reduce to
the nonrelativistic form for small velocities.

We discussed the limitations on the concept of causality imposed by
the uncertainty principle.

We discussed how relativistic probability ampitudes must be a part
of any theory of interaction, including quantum field theory.

\bibliographystyle{apsrev4-1}
\end{document}